\documentclass{article}

\usepackage{arxiv}

\usepackage[utf8]{inputenc} 
\usepackage[T1]{fontenc}    
\usepackage{hyperref}       
\usepackage{url}            
\usepackage{booktabs}       
\usepackage{amsfonts}       
\usepackage{nicefrac}       
\usepackage{microtype}      
\usepackage{lipsum}		
\usepackage{graphicx}
\usepackage{natbib}
\usepackage{doi}
\usepackage[ruled,lined,linesnumbered]{algorithm2e}
\usepackage{bm}
\usepackage{amssymb}
\usepackage{lineno}
\modulolinenumbers[5]
\usepackage{xcolor}
\usepackage{hyperref}
\usepackage{rotating}
\usepackage{booktabs}
\usepackage{lscape}
\usepackage{soul}
\usepackage{xspace}
\usepackage{url}
\usepackage{subcaption}
\usepackage{array}
 \usepackage{float}
 \usepackage{multirow}
\usepackage{colortbl}
\usepackage{graphicx}
\usepackage{mathtools}

\title{FedCSD: A Federated Learning Based Approach for Code-Smell Detection}

\author{
        Sadi Alawadi\\
        Center for Applied Intelligent Systems Research\\
        School of Information Technology\\
        Halmstad University, Halmstad, Sweden\\
        \texttt{sadi.alawadi@hh.se \thanks{Corresponding author.}}
        \And
        Khalid Alkharabsheh\\
        Department of Software Engineering\\
        Prince Abdullah bin Ghazi Faculty of Information and Communication Technology\\ 
        Al-Balqa Applied University,  Jordan\\
        \texttt{khalidkh@bau.edu.jo}
         \And
         Fahed Alkhabbas \\
        Internet of Things and People Research Center\\
        Malm\"{o} University, Sweden\\
        \texttt{fahed.alkhabbas@mau.se}
        \And
         Victor Kebande \\
        Department of Computer Science\\
        Blekinge Institute of Technology, Sweden\\
        \texttt{victor.kebande@bth.se}
        \And
         Feras M. Awaysheh \\
        Institute of Computer Science\\
        Delta Research Centre\\
        University of Tartu, Estonia\\
        \texttt{feras.awaysheh@ut.ee}
        \And
         Fabio Palomba \\
        SeSa Lab \\
        University of Salerno, Italy\\
        \texttt{fpalomba@unisa.it}
        \And
         Mohammed Awad\\
        Arab American University, Palestine\\
        \texttt{mohammed.awad@aaup.edu}
}

\renewcommand{\shorttitle}{FedCSD: A Federated Learning Based Approach for Code-Smell Detection}

\hypersetup{
pdftitle={A template for the arxiv style},
pdfsubject={q-bio.NC, q-bio.QM},
pdfauthor={David S.~Hippocampus, Elias D.~Striatum},
pdfkeywords={First keyword, Second keyword, More},
}

\begin{document}
\maketitle

\begin{abstract}
Software quality is critical, as low quality, or "Code smell," increases technical debt and maintenance costs. There is a timely need for a collaborative model that detects and manages code smells by learning from diverse and distributed data sources while respecting privacy and providing a scalable solution for continuously integrating new patterns and practices in code quality management. However, the current literature is still missing such capabilities. This paper addresses the previous challenges by proposing a Federated Learning Code Smell Detection (FedCSD) approach, specifically targeting "God Class," to enable organizations to train distributed ML models while safeguarding data privacy collaboratively. We conduct experiments using manually validated datasets to detect and analyze code smell scenarios to validate our approach. Experiment 1, a centralized training experiment, revealed varying accuracies across datasets, with dataset two achieving the lowest accuracy (92.30\%) and datasets one and three achieving the highest (98.90\% and 99.5\%, respectively). Experiment 2, focusing on cross-evaluation, showed a significant drop in accuracy (lowest: 63.80\%) when fewer smells were present in the training dataset, reflecting technical debt. Experiment 3 involved splitting the dataset across 10 companies, resulting in a global model accuracy of 98.34\%, comparable to the centralized model's highest accuracy. The application of federated ML techniques demonstrates promising performance improvements in code-smell detection, benefiting both software developers and researchers.

\end{abstract}

\keywords{Software Quality \and Technical Debit \and Federated Learning\and Code Smell Detection}

\maketitle
\section{Introduction\label{sec:Introduction}}

Software quality assurance is a major aspect that occupies the minds of software engineers and the software engineering community at large. 
Consequently, there is a continuous need to maintain the quality of the software, given that it is a determinant in many aspects during and after development. Specifically, software quality assurance determines and detects the software pieces that suffer from low quality in design or programming. These pieces are known as "Code Smells" \cite{fowler1999refactoring}. The existence of code smells does not produce errors during compilation or execution \cite{superiorrefactoring}
, they also  negatively influence the software quality factors \cite{abuhassan2021software, khalidalkharabsheh2018software, brown1998antipatterns}. 

Consequently, the availability of the code smells increases the time and effort required to maintain the software. This extra time and effort is known as technical debt \cite{cunningham1993wycash}, which can be indicated by the presence of code smells. Several terms and concepts have been used to denote code smells, such as antipatterns, disharmonies, design flaws, design defects, code anomalies, design smells, etc \cite{khalidalkharabsheh2018software}. Code smells can be identified in various software components, from instructions to subsystems, and can influence different levels of software granularity, such as methods, classes, and the whole system. Code smell detection is an efficient way to decrease maintenance costs and support the efforts of software developers to improve the quality of software.
Due to the increasing size and complexity of the developed software systems, more automated approaches are needed to improve the activity of code smell detection. 

At present, several approaches concentrate on code smell detection, such as metric/rule-based approaches \cite{choinzon2006detecting,fourati2011metric,marinescu2005iplasma,moha2007decor,munro2005product,shatnawi2015deriving} and machine learning-based approaches \cite{khalidExp, hassaine2010ids, khomh2011bdtex, kreimer2005adaptive, maneerat2011bad, pecorelli2020large, pecorelli2020}. Most of these approaches have been evaluated empirically, and they have obtained  high precision in smell detection. However, there are a set of challenges that constrain their endorsement in the industry, such as the ratio of false negatives and false positives in their findings and the low degree of agreement between them.


To overcome the existing shortcomings and challenges, literature studies, \cite{Shaabyarticle, Azeem-Palomba-She-Whang-2019, fontana2016comparing, khomh2011bdtex, kreimer2005adaptive, maneerat2011bad}, have shown that machine learning-based approaches play a central role in code smell detection and can be more exploited in this direction. Consequently, it is possible to make a quantum leap in improving the detection of the right code smells with high accuracy. ML uses mathematical algorithms to award systems the ability to learn without explicitly programming \cite{Alzubiarticle}.

The use of centralized ML training to detect code smells has been widely investigated. For instance, in \cite{rao2023study,dewangan2022code,dewangan2021novel}, the authors compared the performance of multiple ML algorithms for code smell severity detection over different datasets.
The centralized training process demands a considerable amount of collected and aggregated data, typically in a centralized place such as a data centre, cloud, or server machine. This centralized data aggregation is imperative for constructing an accurate model with quality that adapts to dynamic data, aiming to provide recommendations, decisions, and solutions for specific tasks. However, the considerable expense associated with transferring data to a central hub presents a significant hurdle. Moreover, this data often contains sensitive and private information belonging to the data owner, leading to concerns regarding both data privacy and security. Such matters counter General Data Protection Regulation (GDPR) policies and pose challenges across various sectors, including healthcare, industry, politics, etc.

As a concrete example, let's examine the software industry within the context of our research. In this industry, every company holds a significant stake in understanding the source code and design quality employed by their competitors. Aspects such as performance, maintainability, and reusability are of utmost importance. These companies are keen on leveraging this invaluable data to elevate the quality of their software products. They aim to identify and rectify anomalies and deficiencies in their codebase while refining coding practices and policies to produce top-notch enterprise software. However, it's crucial to note that no company within this competitive landscape will release their private data, including their source code and design details. Instead, they are in pursuit of techniques that allow them to extract valuable insights and knowledge from other companies' data in a secure and non-invasive manner.

In this regard, Federated Learning (FL) emerges as an efficient solution that maintains data privacy and security. FL differs from centralized ML in migrating the ML model to the data's source for training, typically on the edge side \cite{awaysheh2023big}. Unlike centralized ML, FL enables all edge-node models to contribute their knowledge without exposing the raw data (source code or design in our case). By employing FL, software enterprises that are hesitant to share their data can internally train their ML models and then transfer the learned model to a designated entity responsible for maintaining software quality \cite{fed908, zhang2021survey}.

The main contribution of this paper is to propose  Federated Learning Code Smell Detection (FedCSD), which, to the best of our knowledge, is the first approach that exploits FL for code smell detection. Specifically, the God Class smell. We show how FedCSD can be applied in settings where multiple software development companies collaborate to improve the quality of their software development projects without the need to share their code. Further, we discuss how FedCSD can improve the traditional code review activity within software development teams. Finally, we present intensive experiments that show the advantages of applying FedCSD to detect code smells in comparison to traditional centralized ML approaches.

The remainder of this paper is organized as follows. Section \ref{sec:Background} introduces background on code smell detection tools, the role of machine learning in code smell detection, federated learning, and data privacy and attacks. Additionally, it discusses related studies. 
Section \ref{sec:methodology} presents the methodology we applied to design and validate our approach. 
Section \ref{sec:FedCSDApproach} describes the proposed approach. 
Next, Section \ref{sec:validation} analyses and discusses the results. 
while Section \ref{sec:critical} discusses the critical evaluation of the study. Finally, Section \ref{sec:threats} presents the threats of validity and Section \ref{sec:conclusions} presents conclusions and recommendations for future work. 

\section{Background and related work}
\label{sec:Background}

\subsection{Code Smell Detection Tools}
\label{sec:CodeSmellDetection}


Several code smell detection tools have been developed either as standalone or integrated, commercial or open-source, and they support different programming languages and detect various types of code smells Examples of these tools include iPlasma, jCosmo, Incode, DECOR, PMD, Borland Together, and JDeodorant \cite{khalidalkharabsheh2018software, rasool2015review}. However, they have limitations that reduce their effectiveness in industry. Namely, they have a low degree of agreement, lack the capability to analyze software systems implemented in more than one programming language, do not detect a wide set of different types of code smells, lack the interoperability of detection tools with diverse development environments \cite{lewowski2022far}, and how scalable code smell detection techniques are to large-scale codebases.
.


One of the code smells that detection tools focus on the most is Large Class \cite{fowler1999refactoring}, also referred to as God Class \cite{lanza-marinescu-2007}, and the Blob \cite{brown1998antipatterns}.  
In the literature, several studies have focused on detecting the Large Class code smell \cite{khalidalkharabsheh2019improving, kh9634144, kh9585447, ALKHARABSHEH2022, ALKHARABSHEH2022106736, alkharabsheh2016comparacion, alkharabsheh2016grado, counsell2007size, yamashita2015inter}.
 In their systematic mapping study, the authors \cite{khalidalkharabsheh2018software} analyzed close to 400 articles related to code smells and found that Large Class negatively affects different software quality attributes, the most important of which is maintainability.  Based on the above and since Large Class is one of the code smells most frequently detected in software systems, we decided to focus on it in this work.

\subsection{Machine Learning In Code Smell Detection}
\label{sec:MachineLearningandCodeSmellDetection}

In one study  \cite{kreimer2005adaptive}, ML and object-oriented metrics extracted from analyzing software systems were combined into an approach to detect design flaws automatically. The proposed approach was evaluated on three open-source systems. The findings showed that the decision tree effectively detects Large Class and Long Method smells. Another study \cite{maneerat2011bad} utilized ML to predict seven types of design smells (Message Chains, Middle Man, Switch Statement, Long Parameter List, Long Method, Feature Envy, and Lazy Class). The dataset was constructed from a group of 27 metrics gathered from software systems, including design smells.

Furthermore, in \cite{khomh2011bdtex}, the Bayesian Detection Expert (BDTEX) approach was proposed to detect well-known antipatterns named Functional Decomposition, Spaghetti Code, and the Blob. The approach was evaluated on two systems, and the results were compared with the DECOR approach. 
Further, one study \cite{maiga2012support} presented a novel approach that uses the support vector machine and object-oriented metrics to detect Swiss Army Knife, the Blob, Spaghetti Code, and Functional Decomposition antipatterns by analyzing three software systems. The results were compared with the DETEX approach. In another work by \cite{peiris2014towards}, five ML techniques were used based on software metrics to detect different antipatterns. The presented approach was named NiPAD. The study was conducted using one application, and the result showed that the best behavior was obtained by the SVMlinear technique for identifying the One-lane Bridge antipattern. 

A more recent research study by \cite{fontana2016comparing} used 16 ML classifiers to detect Data Class, God Class, Long Method, and Feature Envy code smells. The chosen smells were automatically detected using five tools. The study evaluated 74 software systems, and the results of detection were validated manually by experts in the domain. The findings showed that most of the techniques have a high degree of accuracy. Moreover, in \cite{pecorelli2020}, the performance of metric-based and machine learning-based approaches was empirically compared in terms of code smell detection. The dataset was constructed from 13 software systems and 17 metrics to detect 11 code smells. The results showed that metric-based approaches achieve slightly better performance. Nonetheless, there is a need to conduct more studies on both approaches in order to enhance the precision and efficiency of code smell detection.
Recently, the authors of \cite{ALKHARABSHEH2022106736} conducted a large-scale study that investigated the usefulness of ML techniques for effective design smell detection. The work focused on determining the influence of data balancing on the accuracy of ML techniques during design smell detection. A set of 28 classifiers was used to detect God Class design smells in a dataset of 24 software systems that include 12,587 classes, and the detection results were validated manually by experts. After replicating the experiments on two more datasets, the findings showed there is no significant influence of data balancing on the accuracy of learning classifiers during design smell detection. Moreover, machine learning approaches are efficient in God Class detection. Detecting SQL code smells using code analysis seems like interesting future work \cite{ragab2021depth}.  

All the above studies concluded that standard machine learning-based approaches have a promising and efficient role in the code smell detection context. However, it has some limitations concerning the obtained model. The generated model has been trained on a dataset stored in a centralized place and gathered from different open source software projects located in well-known repositories. In this case, due to the data privacy risks concerning data leaks or misuse and the reluctance of companies to share their complete project data on these repositories, there might be a lack of information about the project context that should be taken into account when training the model, such as architectural patterns, domain-specific requirements, and coding conventions. Therefore, the model may not completely comprehend the intricacies of each software project's coding practises. Consequentially, the model's accuracy will be affected. In this work, to overcome the limitations of previous works, we exploited the advantages of federated learning for code smell detection. On the one hand, our approach involves significant project-specific context information that can be lacking or cannot be shared between companies when training the model, resulting in a more accurate and generalizable code smell detection model. On the other hand, our approach preserves better the data privacy and security of software projects, as companies do not need to share their code repositories.

\subsection{Federated Learning}

Big data systems \cite{awaysheh2021big} and traditional ML approaches are centralized approaches. In general, they require data to be collected and aggregated offline on one site, where the models are trained and deployed \cite{li2020federated, lim2020federated}. These approaches 
have some shortcomings for code smell detection because training and deploying ML models in central nodes requires companies to disclose the source code of their projects. Similarly, distributed learning approaches require the code to be released to the distributed servers. Thus, such approaches do not address the companies’ privacy concerns \cite{cai2019trading,pang2020realizing}. 

To overcome the aforementioned limitations, Google proposed FL, an emerging paradigm that enables users or organizations to jointly train an ML model without releasing their private data \cite{bonawitz2019towards, lim2020federated, mcmahan2017communication}. Fl follows the privacy-by-design philosophy \cite{awaysheh2021security}.  Specifically, in FL settings, companies can train a global ML model to detect code smells collaboratively by aggregating the local trained models’ updated parameters (gradients, weights) and reporting them to the FL server, where the global model will be constructed, then propagating the global model to all the involved companies or contributors. Therefore, by exploiting FL, the companies do not need to share their source code or data, thereby preserving their privacy, and they just share their models’ learned knowledge that collaboratively builds a comprehensive global model to detect code smells.

\subsection{Data Privacy and Security Threats }
Scenarios surrounding traditional ML allude to the fact that centralization plays a major role in holding training data and executing the learning algorithm. In real-world situations, legal restrictions and privacy laws prohibit sharing even well-trained models across diverse participants \cite{truex2019hybrid}. Data being the driving force behind the running of many companies, it is worth noting that tech companies own a majority of ML models; based on how these companies manage the data and code behind the running of operations, this raises pertinent questions of centralization \cite{doku2019towards}. Centralization in this context provides an avenue for a single point of failure and increases the threat levels and potential attack surfaces with the possibility of zero-day vulnerabilities. To accommodate data owners’ constant need for secure and collaborative data execution, an FL environment provides guarantees and assurances from a privacy and security perspective. However, the literature pinpoints other diverse scenarios that emphasize privacy, for example, during model-training \cite{alawadi2021federated} 
and also during security-based model training \cite{kjamilji2021efficient}. 

\begin{table*}[t!]
\centering
\caption{Challenges in Code Smell Detection and Federated Learning opportunities.}
\begin{tabular}{|p{3cm}|p{2.5cm}|p{2.5cm}|p{2.5cm}|p{2.5cm}|}
\hline
\textbf{Challenge in \newline CSD} & \textbf{Traditional Solutions} & \textbf{FL Solution} & \textbf{FL Contribution \newline to CSD} & \textbf{FL Challenges} \\ \hline
Data Privacy and Security & Anonymization (reduces data utility) & Local data processing enhances privacy. & Secure CSD without compromising code. & Ensuring robust local processing handling data diversity. \\ \hline
Diverse and Distributed Data Sources & Centralized data aggregation (size and diversity issues) & Learns effectively from diverse, decentralized sources. & Accurate CSD in varied environments. & Managing data consistency across nodes. \\ \hline
Scalability and Resource Constraints & Powerful centralized servers (costly, less scalable) & Scalable, distributed computational load. & Scalable CSD in large organizations. & Balancing load, ensuring node processing power. \\ \hline
Handling Imbalanced and Sparse Data & Oversampling (introduces biases) & Manages imbalanced data sets by learning from various nodes. & Improved CSD for rare or sparse smells. & Addressing data sparsity and imbalance without biasing. \\ \hline
Integrity and Reliability of Code Quality Assessment & Manual reviews, simplified tools (potential biases) & Decentralized validation for unbiased code quality assessment. & Reliable, unbiased code quality assessment. & Ensuring robustness and impartiality in decentralized validation. \\ \hline
\end{tabular}
\label{table:csd_fl}
\end{table*}

Privacy concerns in code smell detection are important in the software development landscape. This importance is mainly seen when considering the inherent sensitivity of code and data ownership. In the conventional paradigm of centralized code smell detection, where code repositories are extensively analyzed, a critical issue emerges where potential exposure of sensitive or proprietary information is imminent. These repositories often hold the confidential data that forms the backbone of an organization's software projects. Traditional methods inadvertently raise significant privacy concerns when sharing such code repositories or data with external parties. This study assesses these capabilities from the perspective of intentional or unintentional data leakage, which could compromise the trained model \cite{kjamilji2021efficient}.

By harnessing the power of FL, it is envisaged that organizations could collectively train machine learning models while keeping their code and data firmly within their own walls. This collaborative yet privacy-preserving approach ensures that sensitive information remains confidential and proprietary algorithms stay safeguarded.


Security-related threats in the code smell could pose significant challenges to the integrity and reliability of the code quality assessment process. The authors from key assumptions like 
 code injection on original code and adversarial attacks are prevalent in many cases when preparing, training, or deploying learning models. The common form of security-related attacks involves adversarial manipulation, tampering with code and training data during the testing phases, indiscriminate attacks where an adversary makes wrong decisions in order to damage the classifiers, integrity attacks, and availability attacks focused on degrading the usability of the FL system and the code deployed by increasing the positive rate \cite{goodfellow2014explaining,kebande2021active,r2020leveraging}.

 These unauthorized infiltrations can lead to false positives or negatives in code smell detection, rendering the entire process unreliable. In traditional centralized approaches, where code repositories are shared, the risk of such attacks is heightened, as external access to code repositories becomes a potential point of entry for adversaries.

 The suggestions on possible defence strategies that preserve privacy and allow open and closed codes to withstand these attacks are discussed in the subsequent sections of this paper.

\subsection{Why FL in CSD?}

The integration of FL in CSD represents a significant paradigm shift, addressing several limitations inherent in traditional CSD methodologies (See Table \ref{table:csd_fl}). FL's decentralized nature fundamentally enhances data privacy and security, a critical concern in software development where codebases often contain sensitive or proprietary information. By processing data locally at the node level, FL circumvents the need to centralize sensitive code, thus preserving confidentiality while enabling practical code analysis. Moreover, FL's handling of diverse and distributed data sources is particularly advantageous in CSD. Traditional approaches typically rely on centralized data aggregation, which needs to improve with the size and diversity of code repositories. However, FL excels at learning from heterogeneous data sources, offering a more robust and inclusive code quality analysis. This scalability is further beneficial in large-scale projects or organizations where FL distributes computational load across multiple nodes, mitigating resource constraints centralized systems face. 

However, implementing FL in CSD has its challenges (See Table \ref{table:csd_fl}). Ensuring robust local data processing while managing data diversity and consistency across various nodes introduces complexity. 
Another significant challenge is addressing data sparsity and imbalance without introducing biases, especially with non-iid datasets \cite{haller2023handling}. Moreover, when detecting rare or subtle code smells. Despite these challenges, FL's ability to provide real-time, decentralized, and privacy-preserving analysis and its scalability and adaptability to diverse datasets make it an effective and valuable method for CSD. The approach enhances the integrity and reliability of code quality assessment and aligns with the evolving needs of modern, distributed software development practices. Thus, while the path to seamlessly integrating FL in CSD involves navigating certain complexities, the overarching advantages it presents in terms of security, scalability, and comprehensive analysis make it a compelling approach in the realm of code quality management.

\section{Research Methodology}\label{sec:methodology}
To address the challenges highlighted in Section \ref{sec:Background} and  better answer of the needs of the software development communities, we propose the FedCSD approach. For this purpose, we applied in an iterative way the design science research method, as we aimed at devising an innovative approach that solves a practical problem and this is supported by the selected methodology. Specifically, we followed the well-defined guidelines for conducting design science research, which comprises five stages, namely, problem explication, requirements definition, artifact design and development, artifact demonstration, and artifact evaluation\cite{johannesson2014introduction}.

In the \textit{problem explication} stage, we reviewed the literature for code smell detection approaches. We found that no studies investigated the use of FL for code smell detection purposes. Accordingly, we used the Goal-Question-Metric (GQM) approach, which is commonly used by the software engineering community \cite{Runeson2009,Wohlin2012}, to formulate our study goal presented in Table \ref{tble:goal}.

\begin{table}[!t]
	\caption{Goal of this study}
	\normalsize
	\begin{tabular}{l|p{140pt}}
		\textit{Purpose}   & Analyze \\ \hline
  
		\textit{Object}     & set of classes\\  \hline
		\textit{With the purpose of}    & Evaluation\\\hline 
  \textit{With respect to}     & the efficiency of federated learning approach to detect God class code smell in different scenarios\\\hline
  
		\textit{Viewpoint} & from researchers and practitioners point of view \\\hline
  \textit{In the context of} & closed-source software companies
	\end{tabular}
	\label{tble:goal}
\end{table}



In the \textit{requirements definition} stage, we defined one requirement based on the defined goal, namely, to devise an approach that exploits FL to detect code smells cross-organizations. Accordingly, we formulated the following research questions and hypotheses:

\textit{
\begin{description}
\item[RQ1] How can federated learning be effectively leveraged for God Class code smell detection?
\newline \textit{Objective:} by answering this RQ, we aim to understand how FL can be applied within each software development company and also across different companies to detect code smells and consequently improve the software systems' quality. 
\item[RQ2] {How does the use of federated learning affect the quality of the resulting ML model compared to the individual models generated by centralized training approaches? }
\newline{\textit{Objective:} by answering this RQ, we aim at comparing the performances of FL and centralized ML models in detecting code smells.} 
\end{description} 
}

The null hypotheses have been formulated as follows:


\textbf{Hypothesis 1:} \textit{Federated learning cannot be effectively leveraged for God Class code smell detection.\\}
\textbf{Hypothesis 2:} \textit{Federated learning does not improve the quality of the resulting ML model compared to the individual models generated by centralized training approaches.}
In the \textit{artifact design and development} stage, we proposed the first approach that exploits FL to detect code smells during the software development phase (see Section \ref{sec:FedCSDApproach}). Specifically, our approach shows how multiple organizations can collaboratively exploit FL to train ML models without the need to share their code and use the models to improve the qualities of their code. Further, our approach evolves the traditional code review life cycle by integrating the models trained collaboratively.

In the \textit{artifact demonstration and evaluation} stages, we simulated how our approach can be applied in cross-organizational settings and ran experiments that validate its feasibility, respectively (see Section \ref{sec:validation}).

\section{FedCSD Approach}
\label{sec:FedCSDApproach}

This section presents our approach for Federated Learning Code Smell Detection (FedCSD). To the best of our knowledge, FedCSD is the first proposed approach that evolves the traditional code review life cycle by integrating FL to detect code smells during development. Consequently, our approach enables addressing code smells proactively, unlike the majority of existing approaches, which manage technical debt issues reactively. The approach exploits traditional architectures of FL and is mainly based on the FEDn framework.

First, we introduce how the approach supports multiple organizations to employ FL to collaboratively train ML models that can detect code smells with higher accuracy. Then, we describe how the FedCSD evolves the traditional code review activity within the software development life cycle.

\subsection{FedCSD in Action} 
\begin{figure}[!t]
\centering
\includegraphics[width=\linewidth]{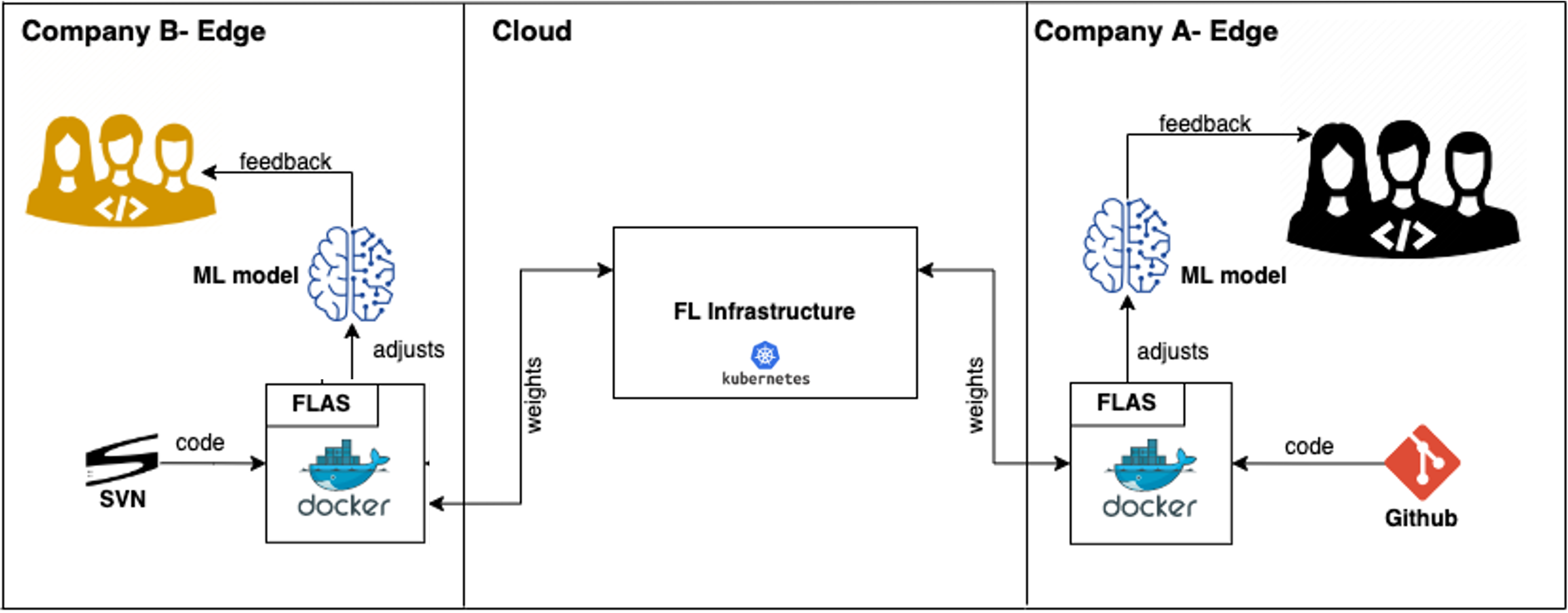}
\caption{Federated Learning in cross-organization settings}
\label{fig:crossfd}
\end{figure}

Figure \ref{fig:crossfd} shows abstractly how FedCSD supports organizations to exploit FL to train ML models and use them to improve the quality of their code. For this purpose, companies need to set up FedCSD containers, which can be realized using Dockers \footnote{\url{https://www.docker.com}}.


For this purpose, companies need to link the containers with both their code repositories and the data pipeline that will be used to train the ML model locally in FL settings. The containers run the client endpoint that automatically starts the local training rounds (i.e., on the edge of the network \cite{awaysheh2022cloud}), reports the updated weights’ resulting from the local training to the FL aggregators, and updates the ML models’ weights according to the results of the global model parameters constructed by the FL aggregator (reducer or server). The FL aggregators (reducers and combiners) run on a shared cloud environment that auto-scales based on the number of involved clients using the Kubernetes \footnote{\url{https://kubernetes.io/}} technology. Algorithm \ref{fedavg} shows the FedAvg algorithm used to derive global weights using the weights generated during local training rounds. The models provide feedback to the developers about their code quality, as described in detail below.

\begin{algorithm}[!t]
\DontPrintSemicolon
\SetAlgoLined
\KwIn{$W_t$}
\KwOut{$M(W_t)$}
\BlankLine
\textbf{Server executes:}\\
initialized \textbf{$W_0$}

\SetKwFunction{Server}{FedAVG}
\SetKwProg{Fn}{Function}{:}{}
\Fn{\Server{$k, W_{t-1}, W_t$}}{

\ForEach{$t\leftarrow 1$ \KwTo $r$}{

$S_t \leftarrow$ (sample a random set of clients)\\
\ForEach{client $k \in S_t$ \textbf{in parallel}}
{
$W_{t+1}^k \leftarrow ClientUpdate(k, W_t,N_l)$\;

$W_{t+1} \leftarrow \sum_{k=1}^{k} \frac{n_k}{n} W_{t+1}^k$\
}
$W_t \leftarrow (W_{t-1} + (W_{t}-W_{t-1})/t)$\
}
\Return $M(W_t)$

}
\caption{FedAvg algorithm, where \textbf{k} is the number of clients, \textbf{r} is the number of rounds, $W_i$ is the local model weights and \textbf{M} is the global model weights}
\label{fedavg}
\end{algorithm}

\begin{figure*}[!ht]
\centering
\includegraphics[width=.6\linewidth]{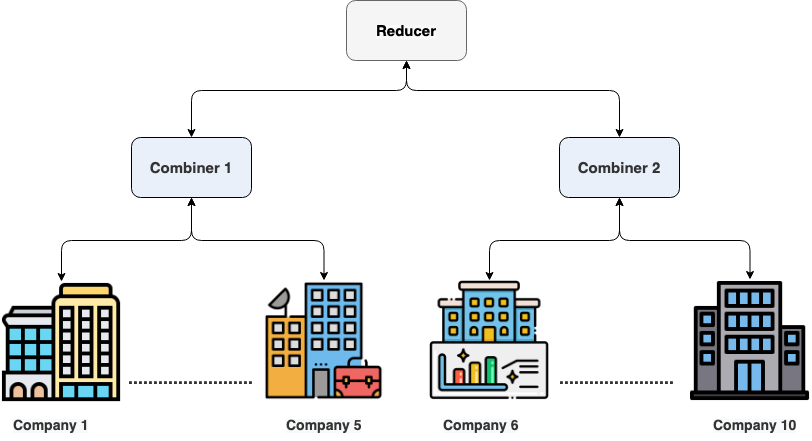}
\caption{A high level representation of FedCSD's main components, including 1 reducer, 2 combiners, and 10 companies}
	\label{fig:Fedn-network}
\end{figure*}

The FL infrastructure of FedCSD adopts the FEDn federated learning framework \cite{ekmefjord2021scalable}. FEDn is an open-source framework that follows the hierarchical MapReduce paradigm. Figure \ref{fig:Fedn-network} illustrates FEDn architecture composed of three layers: reducers, combiners, and clients.

The reducer acts as a server in the server-client paradigm, which has several responsibilities, including the following: (1) monitoring the model training; (2) controlling the communications flow among all the federation components; (3) initiating the seed model with a random weight and then distributing it among the connected combiners; (4) propagating the computing package where the model training and validation instructions are descried to combiners, then from combiners to the connected clients; (5) starting the training (i.e., communication rounds); and (6) aggregating all the updated parameters of the combiners’ local–global models and then averaging them using the FedAvg algorithm (see Algorithm \ref{fedavg}) to construct the final global model. Meanwhile, the combiner represents the intermediate layer responsible for the following: (1) linking the reducer with different client nodes to decrease the reducer’s computation load and the network communication workload; (2) distributing the received model from the reducer across all corresponding clients; and (3) combining all local models’ updated gradients provided by the connected clients using the FedAvg algorithm to build the local–global model. 

Finally, the client layer represents the companies’ local servers (edge nodes), where the data is placed and the local model training rounds are performed. Each client in the federation will receive from the combiner both the ML model and the computing package, which is considered the guideline for the client to train the model. Algorithm \ref{client-update} explains the training process in the client’s local node per communication round. The algorithm returns the model’s updated parameters, which will be reported backward to the upper layer. Each client should be connected to a combiner, but multiple clients can be connected to the same combiner, as can be seen in Figure \ref{fig:Fedn-network}.

\begin{algorithm}[!t]
\DontPrintSemicolon
\SetAlgoLined
\setcounter{AlgoLine}{0}
\KwOut{\textbf{$W_t$}}

\SetKwFunction{client}{ClientUpdate} 
\tcc{Run on client k}
\SetKwProg{Fn}{Function}{:}{}
\Fn{\client{$k, W_t$}}{

$\beta$ $\leftarrow $ (split~$D^k$~into~mini~batches)\\
\For{$local~epoch~e_i \in 1, \dots e$}{
\For{batch~b $\in$ $\beta$}{
$W_t \leftarrow W_t - \eta \nabla l(W_t,b)$\;
}
}
\KwRet{$W_t$}\;
}
\caption{Local client update, where \textbf{k} is the number of clients, \bm{$D^k$} is client k local dataset, \textbf{e} is the number of local epochs, and $\eta$ is the learning rate}
\label{client-update}
\end{algorithm}

\subsection{Evolved Code-Review Activity}

Code review is one of the main activities that reduces maintenance costs and technical debt. Detecting and addressing code smells early prevents them from becoming more complex problems in the future, especially when the software systems are large-scale. Also, developers can improve their knowledge and learning experience by discussing and determining refactoring opportunities suggested by ML models. Also, thanks to FL, our approach enables the generation of ML models that are trained on a variety of software systems from multiple companies without the need to share the code bases of the different projects. 

\begin{figure}[!t]
\centering
\includegraphics[width=\linewidth]{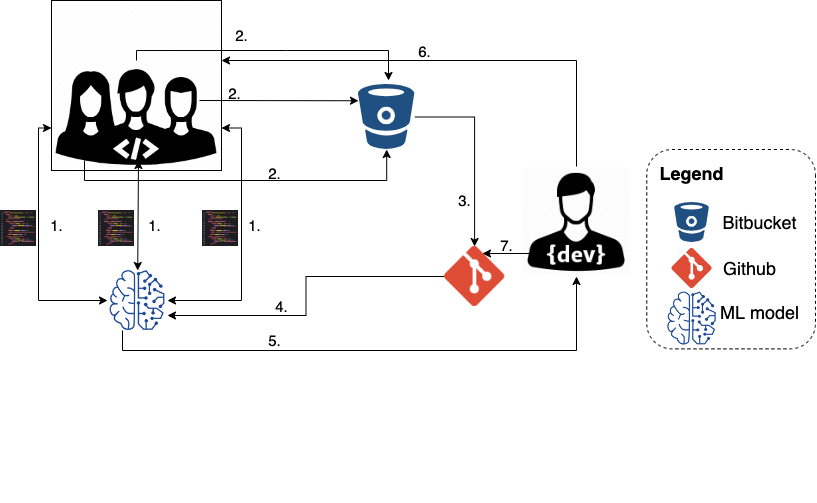}
\caption{The evolved code review cycle}
\label{fig:localsd}
\end{figure}

Figure \ref{fig:localsd} shows the code-review activity, which is part of the traditional software development life cycle, evolved by exploiting FL to improve the code quality. The cycle starts when engineers or developers request to check their individual code for smells before creating code review requests to their team leads or other experienced team members (\textbf{step 1}). Then, the ML model detects smells in the developers’ individual code and provides them with feedback (\textbf{step 2}). Accordingly, the developers update their individual code and then create code review requests (e.g., using Bitbucket) (\textbf{step 3}). After that, the team members responsible for reviewing the different code submitted for review receive feedback from the ML model about smells detected in the entire code being reviewed (\textbf{steps 4 and 5}). Accordingly, the code reviewers provide feedback, possibly to multiple developers, to adjust the code to improve its quality (\textbf{step 6}). This cycle continues until all the comments on the code are addressed and no smells are detected. Consequently, the code submitted for review is approved and merged to the suitable branch (e.g., a release branch on GitHub) (\textbf{step 7}).


\section{Results and Discussion}
\label{sec:validation}

To validate and evaluate the feasibility of the proposed approach in code smell detection activity, we designed the following experiments:
\begin{enumerate}
    \item \textbf{Experiment 1}: Train the code smell detection model for each dataset centrally and evaluate its performance in terms of accuracy.

      \item \textbf{Experiment 2}: Simulate a new coding behavior scenario for a real company by evaluating the trained ML model using parts of other datasets (cross-validation).
      \item \textbf{Experiment 3}: Evaluate our approach by splitting the three datasets into different chunks to simulate 10 distinct companies that will participate in training the global ML model.
  
\end{enumerate}

The code used to run the experiments is available via Github \footnote{\url{https://github.com/saadiabadi/codeSmill.git}}.  

\subsection{Experimental settings}
\label{sec:exps}
In this section, we describe the datasets used in this study. After that, we describe the Long Short-Term Memory (LSTM) algorithm to automatically detect the god class code smell.

\subsubsection{Dataset}
\label{sec:Datasets}
To examine the proposed approach, we used three datasets from the literature: \cite{AlkharabshehCDC19, arcelli2016comparing, pecorelli2020large}. The details of these datasets concerning the number of classes, the number of methods, and the total lines of code in each software are shown in Tables \ref{tab:Pecdataset}, \ref{tab:Fontanadataset}, and \ref{tab:Datacharacteristcs}, respectively, while the comparisons among them are shown in Table \ref{tab:Datasets}. The 1\textsuperscript{st} (\textbf{Pecorelli et al.}) and 2\textsuperscript{nd} (\textbf{Fontana et al.}) datasets were constructed by \cite{pecorelli2020large} and \cite{arcelli2016comparing}, respectively, whereas the 3rd (\textbf{Khalid et al.}) was collected by our team. The set of software systems used in each dataset were open-source, written in Java, came from different domains and size categories, and were available in different repositories, such as Github and SourceForge. Moreover, they are well-known and widely used in the code smell detection context. Concerning (our) dataset (i.e., the 3\textsuperscript{rd} one), we followed concrete criteria to collect the software systems from the repositories, including the number of downloads, availability in several versions, and the history of software systems' maintenance. Due to the huge number of systems that met the criteria, we randomly selected twenty-four systems. The datasets focused on detecting different types of code smells, and the God Class was one of them. According to Fowler \cite{fowler1999refactoring}, a large class is a class that tries to do too many tasks, making it very large regarding the total number of lines of code, number of methods, number of variables, and dependencies with other classes. Therefore, the possibility of duplicate code will increase. Moreover, this class has high complexity as well as low cohesion. \cite{brown1998antipatterns}.
Table  \ref{tab:Datasets} presents the characteristics of the chosen datasets in terms of the number of projects, the number of classes, the number of detection tools, and the number of detected
God Classes (GC) using tools (GC-Tool), the number of human experts who participated in the manual validation process, and the number of God Classes detected by experts (GC-Experts). The total number of software systems was 111 and was formed of more than $80,000$ classes. Each dataset was analyzed automatically by a set of detection tools, and the detection results were manually validated by a group of human experts who have good knowledge of code smell detection. The results of the manual validation were formulated as a binary decision (God Class = 1, Not God Class = 0). As a result, the number of false positives God Classes was reduced in all datasets from $2,696$ to $721$, which represents a $26\%$ reduction. To meet the study's objective, we preprocessed all datasets to have the same features and format. 
Table \ref{tab:features}  reports the 16 features of the dataset, their definitions, and the quality dimensions of different software levels. The replication package in  \cite{Replication} includes all the datasets.

\begin{table}[!htbp]
\scriptsize
 \begin{center}
 \caption{Pecorelli et al. dataset characteristics. \label{tab:Pecdataset}}
\begin{tabular}{|l|r|r|r|} \hline
 {\textbf{Project}}    & {\textbf{NOC}}   & {\textbf{NOM} }  & {\textbf{TLOC}}  \\ \hline \hline
ant-rel-1.8.3	          &	$1,473$ &  $13,213$	     & $119,256$      \\ \hline
 argouml-VERSION\_0\_14          &	$1,373$   &  $9,045$    & $199,075$       \\ \hline
 cassandra-cassandra-1.1.0                  &	$699$    &  $11,360$   & $110,712$       \\  \hline 
 apache-wicket-1.4.11                     &	$1,568$    &  $12,429$    & $174,033$    \\ \hline
derby-10.3.3.0                     & $1,746$	   &  $5,987$   & $535,187$   \\ \hline
 hadoop-release-0.2.0                &	$327$	 &  $2,460$     & $34,662$     \\\hline
 hsqldb-2.2.0                  &	$590$   &  $5,004$    & $254,014$  \\\hline
 incubator-livy-0.6.0-incubating                 &	$1,016$   &  $450$ & $130,696$ \\ \hline
 nutch-release-0.7                     & $532$  &  $3,220$     & $50,578$  \\ \hline
 qpid-0.18               &	$2,172$    &  $21,448$    & $189,271$ \\ \hline
 xerces-Xerces-J\_1\_4\_2               &	$489$  &  $6,088$ & $150,445$  \\ \hline
 eclipse-R3\_4                  &	$5,061$   &  $924$   & $423,423$   \\ \hline
 elasticsearch-v0.19.0                   &	$1,395$   &  $21,739$   & $315,619$    \\ \hline

 \end{tabular}
 \end{center}
  \end{table}


\begin{table}[!htbp]
\scriptsize
 \begin{center}
 \caption{Fontana et al. dataset characteristics. \label{tab:Fontanadataset}}
\begin{tabular}{|l|r|r|r|} \hline
 \textbf{Project}   &  \textbf{NOC}     & \textbf{NOM}& \textbf{TLOC}   \\ \hline \hline
aoi-2.8.1	        &   799            &	688       & 136,533        \\ \hline
argouml-0.34        &    2,361           &  18,015     & 284,934       \\ \hline
axion-1.0.M2        &    223            &   2,989      & 28,583        \\ \hline
castor-1.3.1        &     1,542          &   11,967    & 213,479             \\ \hline
cobertura-1.9.4.1   & 107                 &  3309    &   58,364                 \\ \hline
colt-1.2.0          & 525               &  4,143       & 75,688               \\ \hline   
columba-1.0         & 1,188              &   6818      & 109,035                \\ \hline 
displaytag-1.2      &  128              &  1,064           & 20,892                \\ \hline
drawswf-1.2.9       & 297               &    2,742         & 38,451              \\ \hline
drjava-20100913-r5387     & 225               &  10,364           & 130,132                \\ \hline
emma-2.0.5312       & 262               & 1,805            &   34,404           \\ \hline
exoportal-1.0.2     & 1,855              &  11,709           &  102,803             \\ \hline
findbugs-1.3.9      &1,631               &  10,153           &  146,551             \\ \hline
fitjava-1.0.1       & 60                & 254            &   2,453             \\ \hline
fitlibraryforfitnesse-20100806&795               &  4,165           &   25,691             \\ \hline
freecol-0.10.3      &  1,244             &   8,322          & 163,595                \\ \hline
freecs-1.3.20100406 &  131              &  1,404           &   25,747              \\ \hline
freemind-0.9.0      & 849               &   5,788          &    65,687                 \\ \hline 
galleon-2.3.0       & 764               &  4,305           &  12,072                \\ \hline
ganttproject-2.0.9  & 959               &  5,518           & 58,718                 \\ \hline
heritrix-1.14.4     & 649               &  5,366           &  9,424                  \\ \hline
hsqldb-2.0.0        & 465               &  7,652           &   171,667                 \\ \hline
itext-5.0.3         &  497              & 5,768            &  117,757               \\ \hline
jag-6.1             & 255               & 145            &  24,112                 \\ \hline
jasml-0.10          & 48                & 524            & 6,694                 \\ \hline 
jasperreports-3.7.3 &  1,571             &  17,113           &   260,912                \\ \hline 
javacc-5.0          & 102               & 808            &    19,045               \\ \hline
jedit4.3.2          & 1,037              & 656            &     138,536              \\ \hline
jena-2.6.3          & 1,196              & 99            &   117,117               \\ \hline
jext-5.0            & 485               &  2,169           &  34,855                 \\ \hline
jFin\_DateMath-1.0.1&  58               &  541           &     7,842              \\ \hline
jfreechart-1.0.13   & 960               &   1,181          &    247,421                \\ \hline
jgraph-5.13.0       & 399               & 2,996            &      53,577            \\ \hline
jgraphpad-5.10.0.2  & 426               &  1,879           &    33,431                \\ \hline
jgrapht-0.8.1       & 299               & 1,475            &    28,493            \\ \hline
jgroups-2.10.0      & 1,093              & 8,798            &  126,255                 \\ \hline
jhotdraw-7.5.1      & 968               &  7,232           &      104,357              \\ \hline
jmeter-2.5.1        & 909               &    8,059         &    113,375               \\ \hline
jmoney-0.4.4                 & 190                   & 713      &    9,457                  \\ \hline
jparse-0.96                  & 65                   &  780     &        16,524             \\ \hline
jpf-1.0.2                     &  121                  &  1,271     & 18,172                    \\ \hline
jruby-1.5.2                  &  2,023                  & 17,693       &     199,533                \\ \hline
jspwiki-2.8.4                &   405                 & 2,714      &        69,144             \\ \hline
jsXe-04\_beta                &  100                  &  703     &     1,448                \\ \hline
jung-2.0.1                   & 786                   &   3,884    &   53,617                  \\ \hline
junit-4.1                    & 204                   &  1,031     &  9,065                   \\ \hline
log4j-1.2.16                 & 296                   &  2,118     &  34,617                   \\ \hline   
lucene-3.5.0                 & 1,908                   &  12,486     &      214,819               \\ \hline
marauroa-3.8.1               &  208                  & 1,593      &       26,472              \\ \hline
megamek-0.35.18              &  2,096                  & 13,676      &  315,953                   \\ \hline
mvnforum-1.2.2-ga            & 338                   &  5,983     &      92,696               \\ \hline
nekohtml-1.9.14              & 56                   &  502     &   10,835                  \\ \hline
openjms-0.7.7-beta-1         &  515                  &   379    &     68,929                \\ \hline
oscache-2.4.1                &  66                    &  629     &     11,929                \\ \hline
picocontainer-2.10.2         & 208                   &  1,302     &   12,103                  \\ \hline
pmd-4.2.5                    & 862                   &  5,959     &      71,486               \\ \hline 
poi-3.6                      & 233                   &  19,618     &     299,402                \\ \hline 
pooka-3.0-080505             &  813                  & 68,127      &    68,127                 \\ \hline
proguard-4.5.1               & 604                   &  5,154     &   82,661                  \\ \hline
quartz-1.8.3                 &  280                  & 2,923      &    52,319                 \\ \hline  
quickserver-1.4.7            &  132                  & 1,278      &   18,243                  \\ \hline 
quilt-0.6-a-5                &  66                  &  641     &   8,425                  \\ \hline 
roller-4.0.1                 &  567                  &  5,715     &    78,591               \\ \hline
squirrel\_sql-3.1.2          &  153                  &  689     &          8,378           \\ \hline 
sunflow-0.07.2               & 191                   &   1,447    &     24,319                 \\ \hline
tomcat-7.0.2                 & 1,538                   & 15,627      &     283,829                 \\ \hline
trove-2.1.0                  &  91                  &   585    &   8,432                   \\ \hline
velocity-1.6.4               & 388                   &  2,957     &  5,559                    \\ \hline
wct-1.5.2                    &  606                  & 5,527      &      69,698               \\ \hline
webmail-0.7.10               &  118                  &  1,092     &  14,175                    \\ \hline
Weka-3.7.5                   & 2,045                   &  17,321     &     390,008                \\ \hline
xalan-2.7.1                  &  1,171                  &   10,384    &  312,068                    \\ \hline  
xerces-2.10.0                &    789                &  9,246     &    188,289                 \\ \hline
xmojo-5.0.0                  &    110                &   1,199    &      31,037               \\ \hline


 \end{tabular}
 \end{center}
  \end{table}

\begin{table}[!htbp]
 \begin{center}
\caption{Khalid et al. dataset characteristics. \label{tab:Datacharacteristcs}}
\begin{tabular}{|l|r|r|r|} \hline
{\textbf{Project}}          & {\textbf{NOC} }    & {\textbf{NOM}} & {\textbf{TLOC}}   \\\hline
JCLEC-4-base          &	311       &	1,647 & 37,575       \\ \hline
FullSync-0.10.2       &	169	      & 1,467 & 24,323      \\ \hline
AngryIPScanner-3.0	  &	270	      &  1,228 & 19,965     \\ \hline
SQuirreL-1.2          & 1,138     &	19,031     & 71,626     \\ \hline
Javagraphplan-1.0     &	50        &	537 & 1,049     \\ \hline
DigiExtractor-2.5.2   &	80        &	 523    & 15,668     \\ \hline
JFreechart-1.0.X      &	499       &	  8,024 & 206,559    \\  \hline
Plugfy-0.6	          &	28        &	103 & 2,337     \\ \hline
sMeta-1.0.3     	  &	222       &	1,912 & 30,843     \\ \hline
Ganttproject-2.0.10   &	621       &	5,047 & 66,540   \\ \hline
xena-6.1.0            &	1,975     &	1,272 & 61,526     \\ \hline
pmd-4.3.x             &	800	      & 6,021 & 82,885     \\ \hline
JDistlib-0.3.8        &	78        &	1,027 & 32,081     \\ \hline
Matte-1.7             &	603       &	4,170 & 52,067     \\ \hline
JasperReports-4.7.1   & 1,797     &	 18,781    & 350,690    \\ \hline
Mpxj-4.7              & 553       &	11,634 & 261,971     \\  \hline
Apeiron-2.92          &	62        &	702 & 8,908  \\ \hline
OmegaT-3.1.8          &	716	      & 5,115 & 121,909     \\ \hline
Lucene-3.0.0	      &	606       &	12,459 & 81,611      \\ \hline
KeyStoreExplorer-5.1  &	384       &	 2,535    & 83,144      \\ \hline
Freemind-1.0.1        & 782	      & 6,824     & 106,396    \\ \hline
heckstyle-6.2.0      &	277       & 606 & 41,104      \\  \hline
jAudio-1.0.4	      & 416       &  4,799    & 117,615      \\ \hline
JHotDraw-5.2          &	151       &	1,497 & 17,807    \\
\hline

 \end{tabular}
 \end{center}
\end{table}

\begin{table}[!htbp]

 \scriptsize
\caption{Characteristics of the datasets used in this study}
\label{tab:Datasets}

\scalebox{.8}{
\begin{tabular}{|l|r|r|r|r|r|r|}
\hline
{\textbf{Dataset}} & {\textbf{\#Project}} & {\textbf{\#Class}} & {\textbf{\#Tool}} & {\textbf{\#GC-Tools}} &{\textbf{\#Experts}} & { \textbf{GC-Experts} }\\ \hline \hline 
1st (\textbf{Pecorelli et al.} \cite{pecorelli2020large})& $13$ & $18,441$  & $1$ & $318$ & $2$ & $96$\\ \hline 
2nd (\textbf{Fontana et al.} \cite{arcelli2016comparing})& $74$ &$55,000$ & $2$ &$420$ & $3$ & $140$\\ \hline 
3rd (\textbf{Khalid et al.} \cite{AlkharabshehCDC19})& $24$ & $12,587$& $5$ & $1,958$ & $3$ & $485$  \\ \hline 
\end{tabular}
}
\end{table}

\begin{table} [!ht] 
\scriptsize 
 \begin{center}
\caption{Dataset features. \label{tab:features}}
 \begin{tabular}{|l|l|l|l|l|l|} 
\hline
   { \textbf{No.} }  &  { \textbf{Feature}}   & {\textbf{Definition}}& {\textbf{Granularity}} & {\textbf{Dimension}}  \\ \hline \hline
1 & TLOC   & Total Lines of Code     &  Project  & Size   \\ \hline
2 & NCLOC & Non-Comment Lines of Code  & Project  & Size  \\ \hline
3 & CLOC  & Comment Lines of Code     &  Project & Size  \\ \hline
4 & EXEC  & Executable Statements     &  Project & Complexity  \\ \hline
5 & DC    & Density of Comments      &  Project  & Complexity  \\ \hline
6 & NOT   & Number of Types          &  Package  & Complexity \\ \hline
7 & NOTa  & Number of Abstract Types   & Package & Complexity \\ \hline
8 & NOTc  & Number of Concrete Types   &  Package  & Complexity \\ \hline
9 & NOTe  & Number of Exported Types   &  Package & Complexity \\\hline
10 & RFC   & Response for Class        & Class  & Coupling   \\\hline
11 & WMC   & Weighted Methods per Class & Class  & Complexity\\\hline
12 & DIT   & Depth in Tree              &  Class & Inheritance\\\hline
13 & NOC   & Number of Children in Tree  & Class & Inheritance\\\hline
14 & DIP   & Dependency Inversion Principle  & Class & Coupling\\\hline
15 & LCOM  & Lack of Cohesion of Methods  & Class & Cohesion\\\hline
16 & NOA   & Number of Attributes  & Class & Size\\ \hline
\end{tabular}
 \end{center}
 \end{table}

\subsubsection{Long Short-Term Memory (LSTM) Algorithm}
\label{sec:LSTM}
We used the Long Short-Term Memory (LSTM) model to detect the God Class code smell over the datasets mentioned earlier. The structure of LSTM \cite{sunjaya2023forecasting} depends on three gates: an input gate, a memory and forgetting gate, and an output gate. The input gate regulates the flow of information; the forget gate ensures that unimportant information is forgotten. Ft refers to the following mechanism.
\begin{equation}
\label{eqn:Ft}
\textit{F}_{t} = \sigma(\textit{W}_{f} \cdot [\textit{h}_{t-1},
\textit{x}_{t}] + \textit{b}_{f} )
\end{equation}
The \(\textit{i}_{t}\) represents the input gate that is used to retain the neural network’s state and to determine which data will be incorporated into the cell’s state
\begin{equation}
\label{eqn:it}
\textit{i}_{t} = \sigma(\textit{W}_{i} \cdot [\textit{h}_{t-1},
\textit{x}_{t}] + \textit{b}_{i} )
\end{equation}

The output gate \(\textit{O}_{t}\) presents what extent and how information is filtered out of the neural network.
\begin{equation}
\label{eqn:Ot}
\textit{O}_{t} = \sigma(\textit{W}_{o} \cdot [\textit{h}_{t-1},
\textit{x}_{t}] + \textit{b}_{o} )
\end{equation}

Where \(\textit{ $\sigma$ }\) is the activation function,  \(\textit{W}_{f}\), \(\textit{W}_{i}\) , and \(\textit{W}_{o}\)   are the weights value,  \(\textit{h}_{t-1}\), is output value before ‘t’, \(\textit{x}_{t}\), is input value at ‘t’, and \(\textit{b}_{t}\), \(\textit{b}_{i}\) , and \(\textit{b}_{o}\) are the bias value for the 3 gates.

The model was implemented using the public TensorFlow framework implementation from Keras \footnote{https://github.com/tensorflow/tensorflow}. The model architecture comprises one LSTM input layer with $16$ dimensions, four dense layers with $72$, $50$, $36$, and $28$ units, and a ReLU activation function.  The output layer distinguishes between God Class and Not God Class. For instance, we fixed the model hyper-parameters per communication round to have a $0.001$ initial learning rate with the Adam optimizer. Then, we fixed the batch size to $32$. Finally, we set the maximum number of epochs to $1$. This model has been used in both centralized and collaborative training experiments (FedCSD).

Further, we simulated 10 different companies using the datasets mentioned earlier by using the Pecorelli et al. \cite{pecorelli2020large} dataset, partitioned to five chunks; the Fontana et al. \cite{arcelli2016comparing} dataset, not partitioned; and the Khalid et al. \cite{ALKHARABSHEH2022106736} dataset, partitioned to 4 chunks. All of the previous data chunks represent 10 different heterogeneous companies, as shown in Figure \ref{fig:Fedn-network} in the client layer. Moreover, we relayed our experiment over Swedish National Infrastructure for Computing (SNIC) Science Cloud \cite{8109140} resources, and all instances used in the experiment have 8 Virtual Centralized Processing Units (VCPU), and 16GB RAM.

\subsection{Evaluation Metrics} To evaluate the FedCSD approach, we used the evaluation metrics Accuracy, Loss Function, Kappa, and ROC Area, which are well-known in the literature for evaluating ML in code smell detection. Each metric evaluates the performance of the proposed approach from a different aspect.

\begin{itemize}
    \item Accuracy represents the ratio of correctly classified samples (true positive and true negative). In this study, it is the percentage of classes that are predicted correctly as God Class/Not God Class. However, the accuracy value falls between 0 and 100 and can be computed using Equation \ref{eq:accuracy}. Higher values indicate a more accurate prediction. 

\begin{equation}
 Accuracy= \frac{TP + TN}{TP +TN + FN + FP}*100\%
 \label{eq:accuracy}
\end{equation}




 \item Loss Function is a method to evaluate ML algorithms concerning how well the obtained model is qualified to predict the expected classification results.  
If the predicted results are distant from the actual results, the value of the loss function will be high. This value denotes the errors in the prediction process and can be reduced through learning the loss function. We used the categorical cross-entropy as a loss function, as shown in Equation \ref{eq:loss}.
\begin{equation}
Loss= -\sum_{i=1}^{N} y_i.\log\hat{y}_i
\label{eq:loss}
\end{equation}

where $\hat{y}_i$ is the model prediction for \textit{i-th} pattern, $y_i$ represent the corresponding real value, and \textit{N} is the total number of samples.

\item Cohen Kappa is a test assessing the concordance between the samples that ML algorithms classified and the labelled data.
The values of the Kappa measure range from -1 to 1, where the higher value denotes a strong degree of concordance. The Cohen Kappa can be computed using Equation \ref{eq:kappa}. 

\begin{align}
\kappa &= \frac{P_o - P_e}{1 - P_e}.\label{eq:kappa}
\end{align}

Where $P_o$ represents the samples ratio agreement, and $P_e$ shows the
expected agreement percentage between samples. Moreover, the interpretation of the kappa values is shown in Table~\ref{tab:KappaInterpretation}.






\begin{table}[!htbp]
\centering
\caption{Kappa Values Interpretation.} \label{tab:KappaInterpretation}
\begin{tabular}{|l|l|}
\hline
\textbf{Kappa} & \textbf{Agreement} \\ \hline \hline 
$kappa < 0.20$ & Poor \\ \hline 
$0.21 \leq kappa < 0.40$ & Fair \\ \hline 
$0.41 \leq kappa < 0.60$ & Moderate \\ \hline 
$0.61 \leq kappa < 0.80$ & Substantial \\ \hline 
$0.81 \leq kappa \leq 1.00$ & Almost perfect  \\ 
\hline
\end{tabular}
\end{table} 

\item The ROC-Area, the area under the Response Operating Characteristic (ROC) curve, is a well-known test that is used to evaluate, organize, and visualize the effectiveness of ML algorithms.
It focuses on identifying the relationships between the specificity and sensitivity of learning algorithms. The ROC values range from 0 to 1. The higher ROC value indicates a better learning model. Table ~\ref{tab:ROCInterpretation} presents the interpretation of the test.

\begin{table}[!htbp]
\centering
\caption{ROC area  Interpretation. \label{tab:ROCInterpretation}}
\begin{tabular}{|l|l|}
\hline
\textbf{Value} & \textbf{Interpretation} \\ \hline \hline 
$0.5 < ROC \leq 0.6$ & Fail \\ \hline 
$0.6 < ROC \leq 0.7$ & Poor \\ \hline 
$0.7 < ROC \leq 0.8$ & Fair \\ \hline 
$0.8 < ROC \leq 0.9$ & Good \\ \hline 
$0.9 < ROC \leq$ 1 & Excellent \\ 
\hline
\end{tabular}
\end{table} 
\end{itemize}

 \subsection{Experiment 1: Centralized training}\label{subsec:exp1}
  In this experiment, a centralized ML model was trained over each mentioned dataset to evaluate the performance of the code smell detection model in a traditional company scenario. Table \ref{tab:FedCSDAccuracy} reports the model’s accuracy per dataset. We noticed that the model has achieved high accuracy over all datasets. For instance, training the ML model over both \textbf{Khalid et al.} and \textbf{Pecorelli et al.} datasets has obtained higher performance, with a slight difference related to the number of smells covered by each dataset. Meanwhile, with the \textbf{Fontana et al.}dataset, the model obtained the lowest accuracy (92.30\%), detecting fewer smells than what actually exists. 
 The nature of the dataset, such as software quality, size, and diversity, plays a main role in the model's accuracy. The number of projects used in each dataset as well as the number of classes were different and belonged to different size categories (large, medium, etc.). Therefore, there are differences in the size of the training dataset used to train the model, which directly influence the model's accuracy, as shown in the cases of \textbf{Pecorelli et al.} and \textbf{Khalid et al.}, which were larger than the \textbf{Fontana et al.} dataset. In addition, the set of software projects came from various software domains (application, development, etc.) and statuses (stable, mature, etc.) and were randomly included in the datasets. All these factors influence the model's accuracy and should be taken into account when producing robust and accurate detection models.
 Therefore, we hypothesized that any change in the company’s coding culture or new workers joining the company with different coding behaviours would affect the ML model’s performance and cause technical debt.
 

 \begin{table}[!htbp]
\centering
\caption{The accuracy achieved by the centralized ML model per dataset\label{tab:FedCSDAccuracy}}
\begin{tabular}{|l|l|}
\hline
\textbf{Dataset }& \textbf{Model Accuracy}\\ \hline \hline 
First dataset (\textbf{Pecorelli et al.}) & $98.90\%$ \\ \hline 
Second Dataset (\textbf{Fontana et al.}) & $92.30\%$ \\ \hline 
Third dataset (\textbf{Khalid et al.}) & $99.15\%$ \\ \hline 
\end{tabular}
\end{table}

\subsection{Experiment 2: ML model cross-evaluation} 

Changing the company's coding culture will likely introduce divergences in model performance, leading to concept drift that can adversely influence the model's outputs. To evaluate the resilience and robustness of the centrally trained model against such shifts, we have simulated new changes in companies' coding behaviour that could appear from a new team member or updates in company policies that significantly impact both their internal culture and their products. Therefore, we trained the ML model over one dataset and validated it over the other two datasets (cross-evaluation). Table \ref{tab:accuracy_cross} reports the accuracy achieved by the model trained based on earlier settings. Notably, when the model was trained on either \textbf{Pecorelli et al.} or \textbf{Khalid et al.} and tested on the other, we noticed a small gap in model accuracy compared with the results obtained in experiment 1 (see \ref{subsec:exp1}) for both datasets, and the model has achieved the highest accuracy (96.30\% and 97.00\%, respectively) using both datasets in this context, which refers to the fact that both companies shared the same coding behaviour and culture. 

In contrast, when training or testing the ML model on the \textbf{Fontana et al.} dataset, a notable and significant drop in model accuracy was observed across all cases between 15\% and 30\%. The model obtained the lowest accuracy (63.80\%) compared to the highest accuracy (97.00\%) achieved using other datasets. It is essential to highlight that the \textbf{Fontana et al.} dataset covers different types of smells with further distribution as other datasets. Furthermore, the drift concept, whether in terms of data or model drift, introduces an additional dimension to the context, highlighting the critical steps needed to improve model performance and adaptability in such scenarios.

In conclusion, this experiment clearly shows the significant impact of coding behaviour or culture changes on the smell detection model. By comparing the results obtained from experiments 1 and 2, we observed examples of the model drift concept affecting performance significantly. At the same time, in other cases, the impact was relatively insignificant. To tackle this problem, we propose our FedCSD approach, where the global model is collaboratively trained and built, leveraging contributions from various companies. This approach effectively captures and adapts to changes in a company's culture and its team's coding behaviour.

 \begin{table}[!tbh]
\centering
\caption{Trained LSTM model evaluated over the other two datasets in a centralized fashion. \label{tab:accuracy_cross}}
\setlength{\tabcolsep}{0.1pt}
\begin{tabular}{>{}c >{}c |>{}c >{}c >{}c |}
\cline{3-5}
\multicolumn{2}{c|}{} & \multicolumn{3}{c|}{\textbf{Testing dataset}}  \\ \cline{3-5} 

\multicolumn{2}{c|}{\multirow{-2}{*}{}}  & \multicolumn{1}{c|}{
    \begin{tabular}[c]{@{}c@{}}First dataset \\(\textbf{Pecorelli et al.})\end{tabular}} 
& \multicolumn{1}{c|}{
    \begin{tabular}[c]{@{}c@{}}Second Dataset \\(\textbf{Fontana et al.})\end{tabular}} &
    \begin{tabular}[c]{@{}c@{}}Third dataset \\ (\textbf{Khalid et al.})\end{tabular} \\ \hline
    
\multicolumn{1}{|c|}{\multirow{3}{*}{ \begin{tabular}[c]{@{}l@{}}\textbf{Training}\\ \textbf{dataset}\end{tabular}}} & \begin{tabular}[c]{@{}c@{}}First dataset \\(\textbf{Pecorelli et al.})\end{tabular}  
& \multicolumn{1}{c|}{\cellcolor{blue!25}} & \multicolumn{1}{c|}{63.80\%}  & 96.30\%  \\ \cline{2-5} 

\multicolumn{1}{|c|}{}  & 
    \begin{tabular}[c]{@{}l@{}}Second Dataset \\(\textbf{Fontana et al.})\end{tabular} & \multicolumn{1}{c|}{79.00\%} & \multicolumn{1}{c|}{\cellcolor{blue!25}} & 80.00\% \\ \cline{2-5} 
    
\multicolumn{1}{|c|}{} & 
    \begin{tabular}[c]{@{}l@{}}Third dataset \\(\textbf{Khalid et al.})\end{tabular}& 
    \multicolumn{1}{c|}{97.00\%} & \multicolumn{1}{c|}{71.00\%} & \cellcolor{blue!25} \\ \hline
\end{tabular}
\end{table}


\subsection{Experiment 3: FedCSD evaluation}

We conducted this experiment to answer the research questions presented in Section \ref{sec:goal} as well as to validate our proposed approach (FedCSD) in terms of global model performance (Accuracy and Loss), prediction agreement (Kappa), sensitivity and specificity (ROC value). In this experiment, we simulated 10 companies to participate in the federation by splitting both the \textbf{Pecorelli et al.} and \textbf{Khalid et al.} datasets into chunks that represent five and four companies, respectively, and keeping the \textbf{Fontana et al.} dataset to represent one company. This allowed us to maintain all clients’ heterogeneity and replicate a real scenario. Consequently, the experiment showed the power of federated learning and mitigated the challenges we faced in the previous experiments. Further, we found that this will reduce the computation cost by leveraging the edge nodes (companies) resources to train the model, preserve each company’s data privacy, construct a global model that has a comprehensive knowledge of code smells accumulated from all clients, and reduce the opportunity of having technical debt if new smells appear.

\begin{figure}[!ht]
\centering
\includegraphics[width=\linewidth]{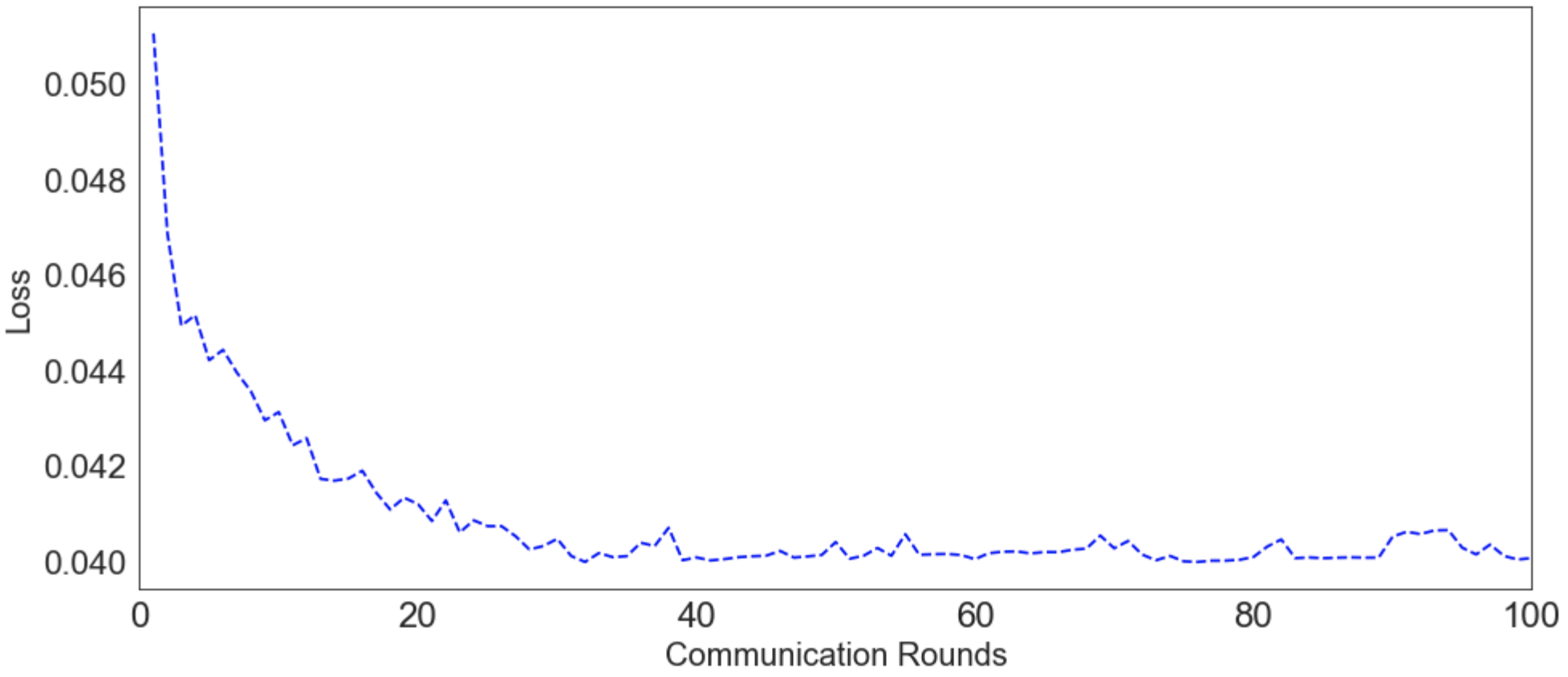}
\caption{Global federated learning model loss function of 10 different clients (companies) for 100 communication rounds }
	\label{fig:Fedn-loss}
\end{figure}

 Figure \ref{fig:Fedn-loss} shows the model’s loss function behavior over the testing set for 100 rounds, which is generally employed over the training and validation sets to optimize the ML algorithm. This metric was calculated using the model prediction for every sample and its corresponding actual output individually, indicating how bad or good the model is. Figure \ref{fig:Fedn-loss} shows the improvement of the model’s learning process after each training round. The testing showed that the model behaved perfectly after round 40, which indicates that the model had reached optimal behavior based on the loss value.  

\begin{figure}[!ht]
\centering
\includegraphics[width=\linewidth]{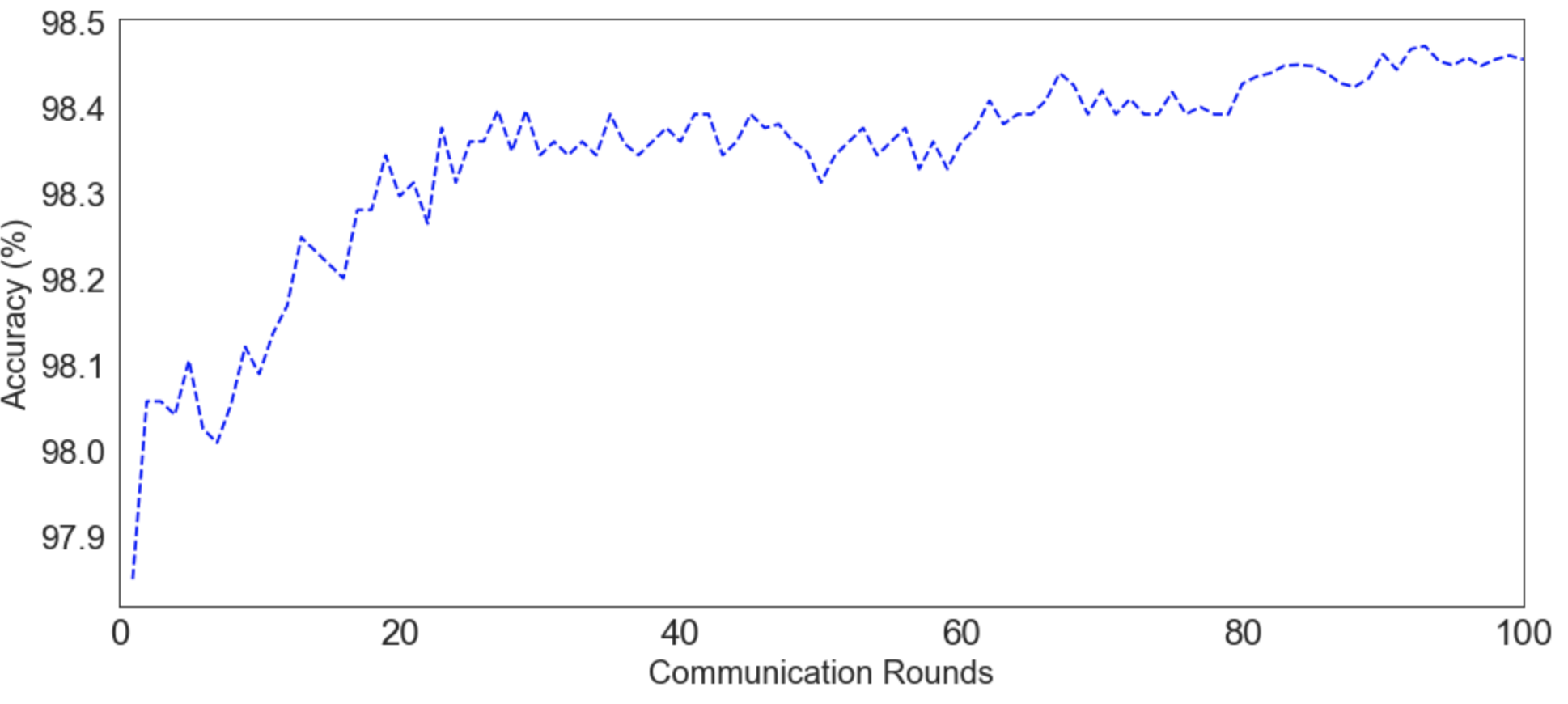}
\caption{Global federated learning model accuracy of 10 different clients (companies) for 100 communication rounds }
	\label{fig:Fedn-acc}
\end{figure}

 After obtaining the optimal optimization of the model using the loss function, the model’s performance was also evaluated in terms of accuracy. Figure \ref{fig:Fedn-acc} illustrates the global model’s performance for 100 training rounds. We noticed that the initial model accuracy was high (97.7\%) and very close to the centralized results. Moreover, the global model’s performance converged in a considerable direction and reached 98.34\%. We noticed that around round 63, the model started to stabilize with only a slight oscillation (0.04\%) until round 95.

 Comparing our FedCSD accuracy with experiments 1 and 2, we argue that our model outperforms both the centralized and cross-evaluation experiments, despite the FedCSD accuracy being a bit lower than the results obtained from the model trained over the Khalid et al. dataset (see Table \ref{tab:FedCSDAccuracy}) by almost 0.1\%. This difference can be ignored in favor of both the model global knowledge and the model stability provided by the FedCSD.

\begin{figure}[!ht]
\centering
\includegraphics[width=\linewidth]{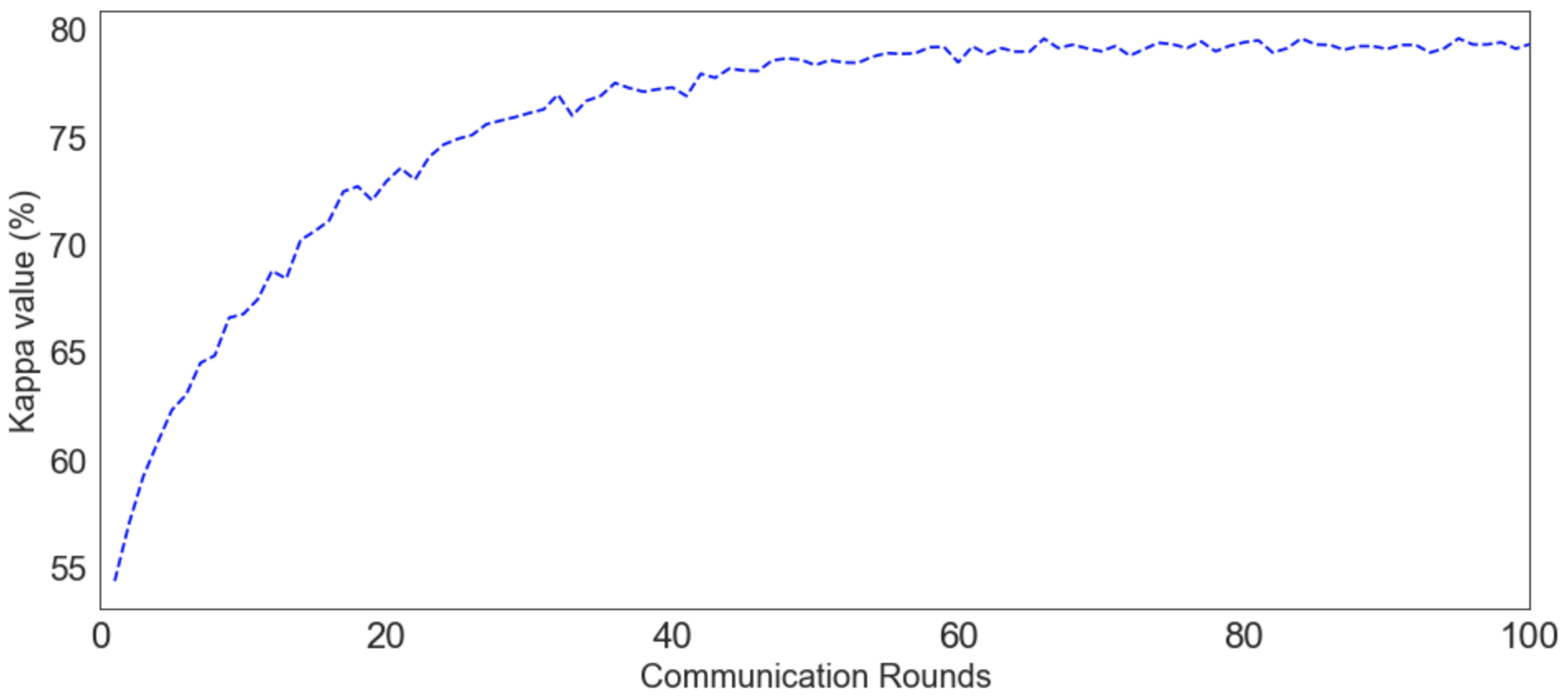}
\caption{Global federated learning model kappa value (\%) of 10 clients (companies) for 100 communication rounds }
	\label{fig:Fedn-kappa}
\end{figure}

Evaluating our FedCSD approach in terms of the agreement ratio between the global model prediction and the corresponding actual value demonstrates the robustness of our approach. Figure \ref{fig:Fedn-kappa} depicts the Cohen Kappa measured while testing the global model for 100 training rounds. There is a significant improvement in the Kappa value per training round, where the initial agreement was low (around 55\%) then linearly converged in the right direction. Further, the Kappa value started to stabilize after round 60 and obtained 79\%, which falls in the substantial range as indicated in Table \ref{tab:KappaInterpretation}. Therefore, the acquired agreement ratio shows the strong learning ability of the FedCSD, which can be generalized for code smell detection problems.

\begin{figure}[!ht]
\centering
\includegraphics[width=\linewidth]{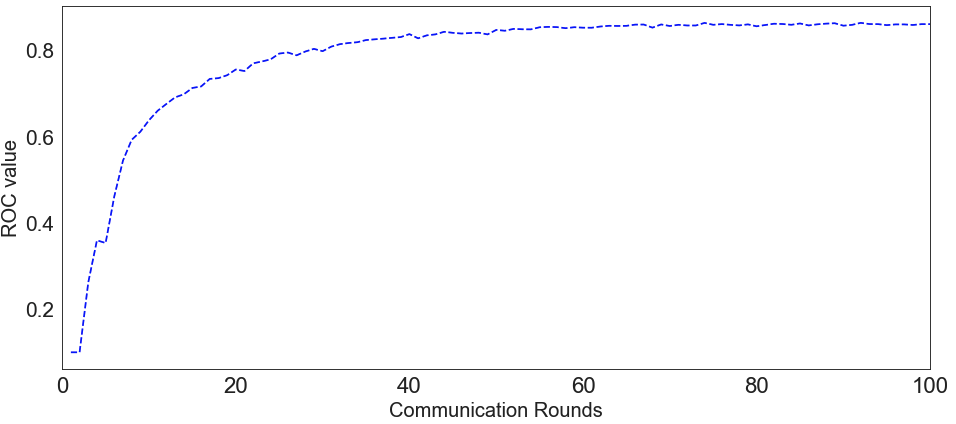}
\caption{Global federated learning model ROC value of 10 clients (companies) for 100 communication rounds }
	\label{fig:Fedn-roc}
\end{figure}

In addition to the previous metrics, we calculated the ROC value for the global model constructed by our approach during 100 rounds. The ROC value is an essential and accurate metric used to evaluate classification problems that do not rely on class distributions. As shown in Figure \ref{fig:Fedn-roc}, the ROC value curve depicts the trade-off between sensitivity (Y-axis) and specificity (X-axis). However, the global model improved linearly per training round, similarly to the previous metrics. Moreover, Figure \ref{fig:Fedn-roc} highlights how after round 60, the model stabilized without any anomalous behavior, which guides us to the conclusion that our approach can capture or learn any new change in the coding culture.


Based on the above, in general, we find that the ML model trained based on the proposed approach (FedCSD) has achieved high performance values according to the loss function, accuracy, kappa, and ROC measurements when detecting code smell. As a result, we conclude that the federated learning approach can effectively be leveraged for God Class code smell detection. Both null hypotheses are therefore rejected

\section{Discussion: Critical Evaluation of the Study}
\label{sec:critical}

The experiments that have been conducted in this study have shown the importance of applying FL to preserve privacy during the training of an ML model, owing to the complexity that is involved during software design. While code smells are seen as perennial issues that affect the quality of software, machine learning, specifically the federated aspect, is touted as a game changer in code smell detection, not only with a higher degree of accuracy but also with precision in preserving essential attributes of the code and data. Accordingly, this study offers an optimal approach aimed at addressing the endemic and perennial privacy issues prevalent during the ML training phase. We are aware of the fact that the knowledge extracted during data training plays a significant role; hence, securely distributing and showing the extracted knowledge from the FL approach emerges as an effective solution to the challenges posed by traditional centralized machine learning methods. Furthermore, leveraging data from open-source software in FedCSD offers considerable advantages, particularly in code smell detection.

 Notably, this approach can be applied in a multitude organizations that will collaboratively be able to not only train ML models but also detect code smells, preserve the privacy of their data, and at the same time expose the relevant security-related risks during the training of ML models. The significant findings of this study can be summarized as follows:

The implications of our findings extend beyond the confines of this research and are addressed as follows:

\begin{itemize}
\item \textbf{Collaborative Software Quality Enhancement:} Our study demonstrates that organisations can effectively harness FL to train and enhance machine learning models collaboratively. This approach holds profound significance for improving the quality of software code while preserving the essential privacy of data and code. Practitioners in software development can leverage this approach to detect and rectify code smells proactively, thus minimizing technical debt and enhancing software maintainability. Therefore, the FedCSD approach introduces a valuable user-feedback component, allowing developers to assess their code quality and detect code smells collaboratively for continuous improvement. This interactive aspect fosters a culture of code quality awareness within development teams, leading to better code practices and enhanced software quality.
 
\item \textbf{Privacy-Preserving Software Development:} The study addresses the critical challenge of privacy in code smell detection, a concern that has often been/not been extensively explored based on existing literature. Our findings emphasize the importance of privacy preservation in machine learning applications and provide a blueprint for other domains where sensitive data is involved.

\item \textbf{On-the-fly Code Smell Detection :} The validity of the experiments that were conducted in this study shows that code detection models aid in not only detecting software design flaws but also improving accuracy during this discourse. The experiments conducted in this study validate the efficacy of code detection models in not only identifying software design flaws but also improving overall accuracy. This holds great promise for practitioners seeking robust tools to assess and enhance their code quality.
\end{itemize}

In addition, we have explored the need for exploring potential security vulnerabilities and adversarial learning threats, which, as a result of implementing FedCSD, may affect integrity and privacy. Firstly, it is pertinent to explore the existence of attack vectors \cite{bouacida2021vulnerabilities}, which have far-reaching implications for the deployment of FedCSD. For example, aspects like model poisoning \cite{zhou2021deep}, where malicious participants may inject erroneous data during training holds. Others include data poisoning \cite{nuding2022data}, model and gradient inversion attacks \cite{liang2023egia}, and membership inference attacks \cite{suri2022subject}. The aforementioned are key adversarial techniques that may defeat the FedCSD implementation. While the scope of this research does not go into an exploration of key mitigation strategies, from a privacy perspective, we identify leveraging differential privacy \cite{galozy2023beyond}, secure aggregation, model encryption, and the use of robust federated optimization algorithms that are essential for ensuring the effectiveness and privacy of federated learning-based approaches such as FedCSD.

 


We argue that leveraging FL to address privacy related aspects in code smell detection is innovative given that, at the time of writing this paper, there was a gap in research on code smell detection using closed datasets, improving accuracy, and involving multiple organizations while also simultaneously preventing code exposure. Importantly, this study outlines the shortcomings of traditional ML, which, from a privacy and security perspective, increases the threat and attack levels of the learning models. While we acknowledge that the study does not go into detail in identifying specific security attacks, it is worth mentioning that we have taken a step in highlighting the generic security-related aspects that could be of interest when deploying FedCSD. Given that the scope of the study is not majorly inclined towards security, we were not concerned with the significance of this aspect during ML phases (as pointed out by \cite{goodfellow2014explaining, kebande2021active, r2020leveraging}). However, we consider this an avenue for future work.

While this study primarily focuses on code smell detection and privacy preservation, it opens up several avenues for future research. Specifically, there is potential for further exploration of security aspects, such as specific security attacks during machine learning phases, as well as the development of more sophisticated privacy-preserving techniques within the Federated Learning framework.

Ultimately, our study bridges the gap in research related to code smell detection using closed datasets, accuracy improvement, and multi-organizational collaboration while maintaining code confidentiality, as was seen in the scenario that was leveraged in this study. We acknowledge the limitations of not delving into specific security attacks, and we recognize this as an area ripe for future investigation. The implications of this research extend to practitioners who seek to elevate software quality while safeguarding data privacy and security, making it not only a significant contribution to academia but also a valuable resource for industry professionals.

Further, we have taken a positive step in acknowledging the previous related studies that have not only laid a firm foundation for this work but have also provided key insights that have significantly consolidated the arguments put forth in this paper.


\section{Threats to Validity}
\label{sec:threats}

This section presents the various threats to the validity of our proposed approach.

\subsection{Construct Validity}
Construct validity concerns the tools and algorithms exploited for code smell detection purposes. Accordingly, one threat concerns the use of the \textit{Fedn} framework. In \cite{ekmefjord2021scalable}, the authors conducted multiple experiments that validated the Fedn framework's scalability, resource utilization, and training accuracy.
Another threat to validity concerns the use of the LSTM algorithm. When we performed the experiments, the Fedn framework supported only deep learning algorithms. We chose the LSTM algorithm because it is known for its ability to store information from previous steps and use that information to influence the output of the current step. Additionally, the LSTM achieved almost the same score as the best code smell detection algorithm reported in \cite{ALKHARABSHEH2022106736}.

\subsection{Internal Validity}

An internal threat to the validity of our approach concerns the distribution of the used datasets with respect to the class instances (i.e., god class/not god class). Unbalanced datasets can affect the quality of the trained ML model. To mitigate this threat, we applied oversampling and undersampling techniques in order to balance the three datasets used in the experiments.

\subsection{External Validity}

An external threat to the validity of our approach concerns the generalization of our experiments' results. Specifically, the application of our approach for detecting code smells, e.g., in commercial and/or non-open-source software systems. To mitigate this threat, we considered three datasets of open-source systems with different application domains and size categories. Indeed, our experiments show that the performance of the ML trained using the datasets outperforms the performance of a centralized ML model trained on an individual dataset.

\section{Conclusions and Future Works}
\label{sec:conclusions}

Great strides have been made in developing federated learning as a distributed AI-based technique as far as the enhancement of privacy of data is concerned. However, at the time of writing this paper, there existed limited or no research that leveraged federated learning in not only detecting code smells but also preserving privacy at the same time. As a result, the research that has been reported in this paper has explored a privacy-aware approach by proposing a Federated Learning Code Smell Detection (FedCSD) that is significant for organizations. The relevance of this proposition is that it enables organizations to ensure software quality and preserve the privacy of their data at the same time by mainly sharing only knowledge as opposed to data. Specifically, in this paper, we demonstrated the application of FL in training ML to detect code smells in different companies' code bases without the need to share those bases. Further, we presented an evolved code review life cycle that integrates our approach. Furthermore, we introduced a variety of datasets that targeted different organizations with code smells, and the outcome showed a higher accuracy not only in the evaluation metrics but also in the global model across all organisations.

The FedCSD global model outperforms the cross-evaluation models, where the FedCSD model was able to detect more smells on a global level which is not detectable individually by the centralized model. Moreover, The FedCSD model shows stability and robustness compared to the results of experiments 1 and 2; even the centralized model of the first and third datasets obtained the highest accuracy, where more resources are required, and there is no data privacy preservation has been considered.

The novelty that backs this study shows a higher relevance when exploring code smells using FL, with dataset two achieving the lowest accuracy of 92.30\% with fewer smells in Experiment 1, while  datasets one and three achieved the highest accuracy with a slight difference of 98.90\% and 99.5\%, respectively. Consequently, in Experiment 2, a significant drop in the model accuracy, lowest accuracy 63.80\% is seen where fewer smells exist in the training dataset. Ultimately, in Experiment 3, where the dataset is split into 10 companies, an accuracy of 98.34\% was achieved by the global model that has been trained using 10 companies for 100 training rounds. In addition, we presented relevant studies that have utilized federated learning in a closely matching context in order to consolidate the key problem and the propositions in this paper. As a result, given that varying datasets have been used, it is the authors opinion that this study outperforms the state-of-the art FL methods. Based on above-mentioned premise, the key objective of this paper, which was identified in the earlier sections, has been reported correctly to best of our knowledge. 

In view of the fore-goings, the authors reiterates that privacy being a perennial challenge among organizations, these propositions gives a guarantee of not only maintaining and preserving privacy but also an assurance of software quality through a FL code smell detection approach. However, owing to the emerging diversification in this area, there are avenues for future work.

In future work, we plan to apply our proposed approach in practice by involving multiple software development companies that develop software systems in different domains. In addition, we plan to extend the approach to detect more types of code smells in software projects implemented in various programming languages. Also, it would be imperative to explore security vulnerabilities, adversarial learning in FedCSD and mitigation strategies.

\bibliographystyle{unsrtnat}

\bibliography{fedbot}
\end{document}